\documentclass{article}
 \usepackage{inchmargs}
 \usepackage{splat}

\newif\ifextended
\extendedfalse

\long\def\extended#1#2{\ifextended #2\else #1\fi}

\newcommand{\rigaddr}{Dept.\ of Mathematical and Computer Sciences\\
 Loyola University\\
 6525 N. Sheridan Rd.\\
 Chicago, IL\ \ 60626-5385}

\newcommand{\leeguanaddr}{Longitude Systems\\
 15000 Conference Center Drive\\
 Chantilly, VA\ \ 20151}

\newcommand{\floor}[1]{\left\lfloor#1\right\rfloor}

\newcommand{\lgN}{\lg N}

\makeatletter

\newcommand{\bibintro}[1]{%
\let\@origthebib=\thebibliography
\let\@origlist=\list
\def\thebibliography{\def\list{#1\par\bigskip\@origlist}\@origthebib}%
}

\makeatother

\newcommand{\inlineciteclosed}
  {\ifx\newblock\undefined\newcommand{\newblock}{\hskip .11em plus .33em minus .07em}\fi}

\inlineciteclosed

\newcounter{oscounter}

\newcommand{\ordinal}[1]{#1\ordsuffix{#1}}

\newcommand{\ordsuffix}[1] 
{\ifcase#1
th\or st\or nd\or rd\or th\or th\or th\or th\or th\or th\or
th\or th\or th\or th\else\setcounter{oscounter}{#1}\ordsuffixrecur\fi}

\newcommand{\ordsuffixrecur}
{\addtocounter{oscounter}{-10}\ifcase\value{oscounter}%
th\or st\or nd\or rd\or th\or th\or th\or th\or th\or th\else
\ordsuffixrecur\fi}

\newcommand{\soda}[1]
    {Proceedings of the \ordinal{#1} Annual {\ifnum#1>3ACM-\fi}SIAM
Symposium on Discrete Algorithms}

\newcommand{\spaa}[1]
    {Proceedings of the
     \ifcase#1\or 1989 \else{\ordinal{#1} Annual }\fi
     ACM Symposium on Parallel Algorithms and Architectures}

 \input{elsart2article}
 \extendedtrue

\begin{document}

\begin{frontmatter}

\title{On the Area of Hypercube Layouts\thanks{Partially supported by NSF grant CCR-9321388.}}

\author{Ronald I. Greenberg\\
 \rigaddr\\
 {\tt http://www.cs.luc.edu/\~{}rig}
 \and
 Lee Guan\\
 \leeguanaddr\\
 {\tt lee\_guan@yahoo.com}
 }

\maketitle

\begin{abstract}
 This paper precisely analyzes the wire density and required area in
standard layout styles for the hypercube.  The most natural, regular
layout of a hypercube of $N^2$ nodes in the plane, in a
$N\times N$ grid arrangement, uses
$\floor{2N/3}+1$ horizontal wiring tracks for each row of
nodes.  (The number of tracks per row can be reduced by 1 with a less
regular design.)  This paper also gives a simple formula for the wire
density at any cut position and a full characterization of all places
where the wire density is maximized (which does not occur at the
bisection).

\end{abstract}

\begin{keyword}
 interconnection networks, hypercube, wire density, VLSI layout area, mincut linear arrangement, optimal linear arrangement, channel routing
 \end{keyword}

\end{frontmatter}

\section{Introduction}
 The (binary) hypercube network has been widely considered as a network
for parallel computing, but its VLSI layout requires a great deal of
wiring area.  Studies of communications capabilities of the hypercube
versus other networks
(e.g.,~\cite{AbrahamP1991,Dally1990,RanadeJ1987,GreenbergG1996ECA,GreenbergG1996ECN})
have varied the width of links between nodes in order to equalize the
hardware costs of the networks being compared under various cost
measures, some of which are closely related to VLSI layout area.

Recall that the interconnection pattern for a hypercube of $N^2$ nodes
can be specified by numbering the nodes from $0$ to $N^2-1$ and
requiring a link between any two nodes whose numbers expressed in
binary differ in exactly one bit.  When the numbers differ in the
$i$th bit from the right, we refer to the link between the nodes as a
dimension $i$ link.  (Though the links between nodes are generally
considered to be bidirectional, we count them as one wire for
simplicity.  Results quoted in this paper must be multiplied by $2$ to
obtain exact correspondence with results given in
Dally~\cite{Dally1990} or Ranade and Johnsson~\cite{RanadeJ1987}.)

\begin{figure}
 \centering
 \input{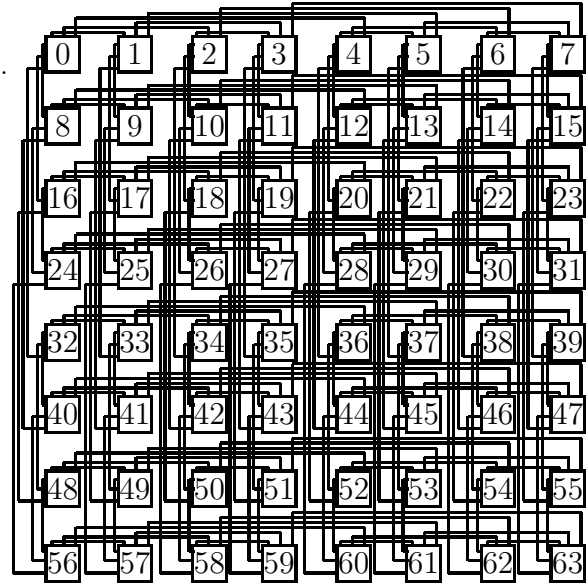}
 \caption{The normal hypercube layout and a naive track assignment for $N^2=64$.}
 \label{fig:normal-layout}
 \end{figure}

The network cost measure used by Dally~\cite{Dally1990} is bisection
width (the minimum number of wires that must be cut to divide the set
of nodes into two equal halves with no connections between them).
This measure may be justified by Thompson's lower
bound~\cite{Thompson1979,Thompson1980} indicating that area is at
least $1/4$ of the square of the bisection width.  Thompson's bound,
however, does not give a precise correspondence between bisection
width and area.  Furthermore, as Dally notes, the maximum wire density
(number of wires that must cross a cutline) does not occur at the
bisection in the ``normal layout'' of the hypercube (nodes placed as
in Figure~\ref{fig:normal-layout}).  (Note that each row and column of
the layout is itself a hypercube, so we can focus henceforth on the
layout of an $N$-node hypercube in a single row.)

Ranade and Johnsson~\cite{RanadeJ1987} consider the actual area required
for the normal layout by bounding the number of horizontal tracks per
row required to lay out the interconnections (following the common
approach of placing vertical wires in one chip layer and horizontal
wires in another).  (The situation involving vertical tracks is
completely analogous to that involving horizontal tracks.)  They
focus, however, on optimality to within an unspecified constant factor
and only upper bound the number of tracks per row as $N-1$, as
obtained by the assignment of wires to tracks illustrated in
Figure~\ref{fig:normal-layout}.

A more sophisticated track assignment by Chen, Agrawal, and
Burke~\cite{ChenAB1993} (with a different ordering of the nodes), yields
$N-\lgN$ tracks per row.\footnote{ We use $\lg x$ for
$\log_{2}x$, and we assume $N$ is a power of $2$.}

A still better measure for the number of tracks per row, utilized
in~\cite{GreenbergG1996ECA,GreenbergG1996ECN}, is
$\floor{2N/3}$.  That this number represents the congestion for
the natural embedding of the hypercube into a square grid also follows
from an independent statement of Nakano~\cite{Nakano1994} and an
argument of Bezrukov \etal~\cite{BezrukovCHRS2000}.

This paper gives a short alternative proof of the congestion result
that also yields a concise formula for the wire density at every cut
position and a full characterization of all positions where density is
maximized.  The analysis is then extended to account for the exact
placement of the terminals and wires in the layout.  It would be
desirable to make all nodes identical, e.g., by placing the
connections of each node in order of dimension (as in
Figure~\ref{fig:normal-layout}); this would be particularly convenient
when implementing the common form of hypercube algorithm referred to
as a ``normal algorithm'' (e.g., see~\cite{Leighton1992}), in which
only one dimension of communication links is used at any step, and the
dimensions are used consecutively.  Uniformity of nodes is also
helpful for assembling the system and for replacing defective nodes.
We show that such a uniform approach incurs a penalty of exactly one
track per row in the VLSI layout, whereas full freedom to permute the
terminals allows a layout with $\floor{2N/3}$ tracks per row.

The rest of this paper is organized as follows.
Section~\ref{sec:density} introduces notation and provides background
regarding the congestion result.  Section~\ref{sec:tracks}, gives a
simple formula for the wire density at each intercolumn position and a
full characterization of those positions where the density is
maximized.  Then the analysis is extended to include the density at
cutlines that run through nodes, which completes the analysis of the
number of wiring tracks required. Section~\ref{sec:alt}, comments on
hypercube layouts in which the nodes are placed differently than in
the normal scheme illustrated in Figure~\ref{fig:normal-layout}.

 \label{sec:intro}

\section{Background}
 As a first step towards determining the usage of wiring tracks in the
normal hypercube layout, we focus on the intercolumn wire density per
row.  We \extended{sketch}{give} here a short \extended{proof, shown
fully in~\cite{GreenbergG2001},}{proof} that the maximum intercolumn
wire density per row in the normal hypercube layout is $\floor{2N/3}$
and that the leftmost intercolumn position where this maximum is
realized is position $\floor{(N+1)/3}$.  In the process we introduce
notation for our main results in the next section and note important
symmetry properties.

We define $f(i,k)$ to be the number of dimension $k$ links (i.e.,
links spanning $2^{k-1}$ columns) that cross intercolumn position $i$
in the normal layout.  Using 0 to denote the position to the left of
{\it all\/} the nodes, it is easy to see that the pattern for
$f(0,k)$, $f(1,k)$, \ldots,~$f(N-1,k)$ is $0$, $1$, $2$,
\ldots,~$2^{k-1}-1$, $2^{k-1}$, $2^{k-1}-1$, $2^{k-1}-2$, \ldots,~$1$,
and repeat as necessary; we may express this as
 \begin{equation}
 f(i,k) =
 i\left(1-2\left(\floor{\frac{i-1}{2^{k-1}}}\bmod2\right)\right)\bmod2^{k} \ .
 \label{eqn:f-i-k}
 \end{equation}
 Then we define $S(i,N)$ to be the total number of connections
crossing intercolumn position $i$ in the normal layout, i.e.,
 \begin{equation}
 S(i,N) = \sum_{k=1}^{\lgN}f(i,k) \ .
 \label{eqn:S-i-n}
 \end{equation}

For the \extended{proof sketch}{proof} in this section, there is also
a more convenient mathematical expression for the maximum intercolumn
wire density and and the leftmost position where the maximum is
realized:
 \begin{eqnarray}
 m(N) &=& (4N-(-1)^{\lgN}-3)/6 \label{eqn:mN} \\
 p(N) &=& (N-(-1)^{\lgN})/3 \label{eqn:pN}
 \end{eqnarray}

Then the result discussed in this section is that
 $\max_{0<i<N}S(i,N) = m(N)$
 and that $i=p(N)$ is the least $i$ at which the maximum is achieved.
 The result follows from the following Lemma and two Theorems:

\begin{lem}
 \label{lem:symmetry}
 $S(i,N) = S(N-i,N)$ for $0<i<N$.
 \end{lem}

\begin{pf*}{\extended{Proof sketch}{Proof}}
 The result follows from showing $f(i,k)=f(N-i,k)$ for
$0<i<N$ and $1\leq k\leq\lgN$, which follows from
\extended{Equation~\ref{eqn:f-i-k}.}{Equation~\ref{eqn:f-i-k}:
 \begin{eqnarray*}
 f(N-i,k) & =
 & (N-i)
   \left(1-2\left(\floor{\frac{N-i-1}{2^{k-1}}}\bmod2\right)\right)\bmod2^{k} \\
 & = & (-i)
   \left(1-2\left(\floor{\frac{-i-1}{2^{k-1}}}\bmod2\right)\right)\bmod2^{k}
   \\ & & \mbox{since $N$ is a multiple of $2^{k}$} \\
 & = & (i)
   \left(1-2\left(\floor{\frac{i-1}{2^{k-1}}}\bmod2\right)\right)\bmod2^{k}
   \\ & & \mbox{since \extended{floor}{the floor} switches parity unless
                $i\equiv-i\equiv2^{k-1}\pmod{2^{k}}$} \\
 & = & f(i,k) \ .  \qed
 \end{eqnarray*}
 }
 \end{pf*}

\begin{thm}
 $S(p(N),N)=m(N)$.
 \end{thm}

\begin{pf*}{\extended{Proof sketch}{Proof}}
 The proof is by \extended{induction on $\lgN$ using
Equations~\ref{eqn:f-i-k}--\ref{eqn:pN} and Lemma~\ref{lem:symmetry}.
\qed}{induction.  The statement is trivial for $N=1$.  The induction
hypothesis is that $S(p(x),x)=m(x)$ for all $x$ that are even powers
of 2 less than $N$ and some $N\geq2$.  From this hypothesis, we
proceed to show that $S(p(N),N)=m(N)$:
 \begin{eqnarray*}
 S(p(N),N) & =
     & f(p(N),\lgN)+ S(p(N),N/2) \mbox{ by Equation~\ref{eqn:S-i-n}} \\
 & = & f(p(N),\lgN)+S(p(N/2),N/2)
       \mbox{ by Lemma~\ref{lem:symmetry} \extended{\&}{and} Equation~\ref{eqn:pN}} \\
 & = & p(N) + m(N/2)
       \mbox{ by Equations~\ref{eqn:pN} \extended{\&}{and}~\ref{eqn:f-i-k} \extended{\&}{and the}~induction hypothesis} \\
 & = & m(N) \mbox{ by Equations~\ref{eqn:mN} and~\ref{eqn:pN}} \ .
       \qed
 \end{eqnarray*}
 }
 \end{pf*}

Now we need only that $S(i,N)\leq m(N)$, but the following theorem
includes additional information to make the proof easier:

\begin{thm}
 $S(i,N) \leq \min\{m(N),m(N)-(p(N)-i)\}$ for $0<i<N$.
 \end{thm}

\begin{pf*}{\extended{Proof sketch}{Proof}}
 We again use induction \extended{on $\lgN$}{and show that the statement follows under the
assumption that it holds for smaller values of $N$}.

\extended{We first use Equations~\ref{eqn:f-i-k}--\ref{eqn:pN} to show
that $S(i,N) \leq m(N)-(p(N)-i)$.  Then we show $S(i,N) \leq m(N)$
by considering the three cases of $i>N/2$, $i \leq p(N)$, and
$p(N) < i \leq N/2$ and using Lemma~\ref{lem:symmetry} again.  \qed}
{
We note first that
 \begin{eqnarray}
 S(i,N) & = & f(i,\lgN) + S(i,N/2) \nonumber \\
   & \leq & i + m(N/2)
            \mbox{ by Equation~\ref{eqn:f-i-k} and the induction hypothesis}
            \nonumber \\
   & = & m(N) - (p(N)-i)
         \mbox{ by Equations~\ref{eqn:mN} and~\ref{eqn:pN}}
         \label{ineq:small-i}
 \end{eqnarray}

All that remains is to show that $S(i,N)\leq m(N)$, which we split
into three cases according to the value of $i$:
 \begin{description}
 \item{\underline{Case I: $i>N/2$}}.  Since
$S(i,N)=S(N-i,N)$ by Lemma~\ref{lem:symmetry}, it suffices to
consider cases II and III.
 \item{\underline{Case II: $i\leq p(N)$}}.  The result follows from
Inequality~\ref{ineq:small-i}.
 \item{\underline{Case III: $p(N)<i\leq N/2$}}.  We have
 \begin{eqnarray*}
 S(i,N) & = & f(i,\lgN) + S(i,N/2) \\
   & = & i + S(\sqrt{N/2}-i,N/2)
         \mbox{ by Equation~\ref{eqn:f-i-k} and Lemma~\ref{lem:symmetry}} \\
   & \leq & i + m(N/2)-(p(N/2)-(\sqrt{N/2}-i))
            \mbox{ \extended{by}{by the} induction hypothesis} \\
   & = & m(N)
         \mbox{ by utilizing Equations~\ref{eqn:mN} and~\ref{eqn:pN}}
         \qed
 \end{eqnarray*}
 \end{description}
}
 \end{pf*}

 \label{sec:density}

\section{Number of wiring tracks}
\begin{figure}
 \centering
 \setlength{\unitlength}{0.00083300in}%
\begingroup\makeatletter\ifx\SetFigFont\undefined
\def\x#1#2#3#4#5#6#7\relax{\def\x{#1#2#3#4#5#6}}%
\expandafter\x\fmtname xxxxxx\relax \def\y{splain}%
\ifx\x\y   
\gdef\SetFigFont#1#2#3{%
  \ifnum #1<17\tiny\else \ifnum #1<20\small\else
  \ifnum #1<24\normalsize\else \ifnum #1<29\large\else
  \ifnum #1<34\Large\else \ifnum #1<41\LARGE\else
     \huge\fi\fi\fi\fi\fi\fi
  \csname #3\endcsname}%
\else
\gdef\SetFigFont#1#2#3{\begingroup
  \count@#1\relax \ifnum 25<\count@\count@25\fi
  \def\x{\endgroup\@setsize\SetFigFont{#2pt}}%
  \expandafter\x
    \csname \romannumeral\the\count@ pt\expandafter\endcsname
    \csname @\romannumeral\the\count@ pt\endcsname
  \csname #3\endcsname}%
\fi
\fi\endgroup
\begin{picture}(3639,499)(139,242)
\thicklines
\put(151,254){\makebox(6.6667,10.0000){\SetFigFont{10}{12}{rm}.}}
\put(1051,479){\line( 0, 1){ 50}}
\put(1051,529){\line( 1, 0){825}}
\put(1876,529){\line( 0,-1){ 50}}
\put(1951,479){\line( 0, 1){ 50}}
\put(1951,529){\line( 1, 0){1800}}
\put(3751,529){\line( 0,-1){ 50}}
\put(1771,269){\framebox(195,210){}}
\put(2236,269){\framebox(195,210){}}
\put(2671,269){\framebox(195,210){}}
\put(3136,269){\framebox(195,210){}}
\put(3571,269){\framebox(195,210){}}
\put(871,269){\framebox(195,210){}}
\put(421,269){\framebox(195,210){}}
\put(1336,269){\framebox(195,210){}}
\put(601,479){\line( 0, 1){ 50}}
\put(601,529){\line( 1, 0){300}}
\put(901,529){\line( 0,-1){ 50}}
\put(976,479){\line( 0, 1){200}}
\put(976,679){\line( 1, 0){1800}}
\put(2776,679){\line( 0,-1){200}}
\put(1501,479){\line( 0, 1){100}}
\put(1501,579){\line( 1, 0){300}}
\put(1801,579){\line( 0,-1){100}}
\put(526,479){\line( 0, 1){100}}
\put(526,579){\line( 1, 0){825}}
\put(1351,579){\line( 0,-1){100}}
\put(2401,479){\line( 0, 1){100}}
\put(2401,579){\line( 1, 0){300}}
\put(2701,579){\line( 0,-1){100}}
\put(451,479){\line( 0, 1){150}}
\put(451,629){\line( 1, 0){1800}}
\put(2251,629){\line( 0,-1){150}}
\put(3301,479){\line( 0, 1){150}}
\put(3301,629){\line( 1, 0){315}}
\put(3616,629){\line( 0,-1){150}}
\put(2326,479){\line( 0, 1){150}}
\put(2326,629){\line( 1, 0){840}}
\put(3166,629){\line( 0,-1){150}}
\put(2851,479){\line( 0, 1){100}}
\put(2851,579){\line( 1, 0){825}}
\put(3676,579){\line( 0,-1){100}}
\put(1426,479){\line( 0, 1){250}}
\put(1426,729){\line( 1, 0){1800}}
\put(3226,729){\line( 0,-1){250}}
\put(518,314){\makebox(0,0)[b]{\smash{\SetFigFont{12}{14.4}{rm}0}}}
\put(968,314){\makebox(0,0)[b]{\smash{\SetFigFont{12}{14.4}{rm}1}}}
\put(1433,314){\makebox(0,0)[b]{\smash{\SetFigFont{12}{14.4}{rm}2}}}
\put(1868,314){\makebox(0,0)[b]{\smash{\SetFigFont{12}{14.4}{rm}3}}}
\put(2333,314){\makebox(0,0)[b]{\smash{\SetFigFont{12}{14.4}{rm}4}}}
\put(2768,314){\makebox(0,0)[b]{\smash{\SetFigFont{12}{14.4}{rm}5}}}
\put(3233,314){\makebox(0,0)[b]{\smash{\SetFigFont{12}{14.4}{rm}6}}}
\put(3668,314){\makebox(0,0)[b]{\smash{\SetFigFont{12}{14.4}{rm}7}}}
\end{picture}
 \caption{Wiring a row in $m(N)=5$ tracks for $N=8$.}
 \label{fig:normal-float}
 \end{figure}
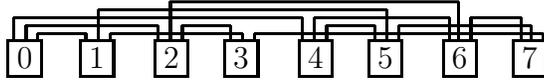

 Though we know the maximum intercolumn wire density per row in the
layout of Figure~\ref{fig:normal-layout}, we still need to determine
the number of horizontal wiring tracks required to route the wires.
Fortunately, an early channel routing algorithm of Hashimoto and
Stevens~\cite{HashimotoS1971}, the left-edge algorithm, guarantees
that the density and number of tracks are equal, since we have no
vertical constraints (e.g., see~\cite{Lengauer1990}).  To obtain a
layout using exactly $m(N)$ tracks, however, we must be free to
permute the locations of connections on each hypercube node so that
the density (maximum number of wires crossing a vertical line) is no
higher when the cutline runs through nodes than when it runs between
nodes.  A layout using $m(N)=5$ tracks for one row of the 64-node
hypercube is illustrated in Figure~\ref{fig:normal-float}.  (This
figure uses a track assignment slightly different than the assignment
produced by the left edge algorithm in order to reduce the number of
wire crossings.)

\begin{figure}
 \centering
 \setlength{\unitlength}{0.00083300in}%
\begingroup\makeatletter\ifx\SetFigFont\undefined
\def\x#1#2#3#4#5#6#7\relax{\def\x{#1#2#3#4#5#6}}%
\expandafter\x\fmtname xxxxxx\relax \def\y{splain}%
\ifx\x\y   
\gdef\SetFigFont#1#2#3{%
  \ifnum #1<17\tiny\else \ifnum #1<20\small\else
  \ifnum #1<24\normalsize\else \ifnum #1<29\large\else
  \ifnum #1<34\Large\else \ifnum #1<41\LARGE\else
     \huge\fi\fi\fi\fi\fi\fi
  \csname #3\endcsname}%
\else
\gdef\SetFigFont#1#2#3{\begingroup
  \count@#1\relax \ifnum 25<\count@\count@25\fi
  \def\x{\endgroup\@setsize\SetFigFont{#2pt}}%
  \expandafter\x
    \csname \romannumeral\the\count@ pt\expandafter\endcsname
    \csname @\romannumeral\the\count@ pt\endcsname
  \csname #3\endcsname}%
\fi
\fi\endgroup
\begin{picture}(3639,549)(139,242)
\thicklines
\put(151,254){\makebox(6.6667,10.0000){\SetFigFont{10}{12}{rm}.}}
\put(451,479){\line( 0, 1){ 50}}
\put(451,529){\line( 1, 0){450}}
\put(901,529){\line( 0,-1){ 50}}
\put(2251,479){\line( 0, 1){250}}
\put(2251,729){\line( 1, 0){450}}
\put(2701,729){\line( 0,-1){250}}
\put(3166,479){\line( 0, 1){200}}
\put(3166,679){\line( 1, 0){450}}
\put(3616,679){\line( 0,-1){200}}
\put(2326,479){\line( 0, 1){300}}
\put(2326,779){\line( 1, 0){900}}
\put(3226,779){\line( 0,-1){300}}
\put(976,479){\line( 0, 1){ 50}}
\put(976,529){\line( 1, 0){900}}
\put(1876,529){\line( 0,-1){ 50}}
\put(2776,479){\line( 0, 1){150}}
\put(2776,629){\line( 1, 0){900}}
\put(3676,629){\line( 0,-1){150}}
\put(1501,479){\line( 0, 1){100}}
\put(1501,579){\line( 1, 0){1800}}
\put(3301,579){\line( 0,-1){100}}
\put(1951,479){\line( 0, 1){ 50}}
\put(1951,529){\line( 1, 0){1800}}
\put(3751,529){\line( 0,-1){ 50}}
\put(1771,269){\framebox(195,210){}}
\put(2236,269){\framebox(195,210){}}
\put(2671,269){\framebox(195,210){}}
\put(3136,269){\framebox(195,210){}}
\put(3571,269){\framebox(195,210){}}
\put(871,269){\framebox(195,210){}}
\put(421,269){\framebox(195,210){}}
\put(1336,269){\framebox(195,210){}}
\put(526,479){\line( 0, 1){100}}
\put(526,579){\line( 1, 0){900}}
\put(1426,579){\line( 0,-1){100}}
\put(601,479){\line( 0, 1){150}}
\put(601,629){\line( 1, 0){1800}}
\put(2401,629){\line( 0,-1){150}}
\put(1051,479){\line( 0, 1){200}}
\put(1051,679){\line( 1, 0){1800}}
\put(2851,679){\line( 0,-1){200}}
\put(1351,479){\line( 0, 1){250}}
\put(1351,729){\line( 1, 0){450}}
\put(1801,729){\line( 0,-1){250}}
\put(518,314){\makebox(0,0)[b]{\smash{\SetFigFont{12}{14.4}{rm}0}}}
\put(968,314){\makebox(0,0)[b]{\smash{\SetFigFont{12}{14.4}{rm}1}}}
\put(1433,314){\makebox(0,0)[b]{\smash{\SetFigFont{12}{14.4}{rm}2}}}
\put(1868,314){\makebox(0,0)[b]{\smash{\SetFigFont{12}{14.4}{rm}3}}}
\put(2333,314){\makebox(0,0)[b]{\smash{\SetFigFont{12}{14.4}{rm}4}}}
\put(2768,314){\makebox(0,0)[b]{\smash{\SetFigFont{12}{14.4}{rm}5}}}
\put(3233,314){\makebox(0,0)[b]{\smash{\SetFigFont{12}{14.4}{rm}6}}}
\put(3668,314){\makebox(0,0)[b]{\smash{\SetFigFont{12}{14.4}{rm}7}}}
\end{picture}
 \caption{Wiring a row requires $m(N)+1=6$ tracks for $N=8$ when the
wires leaving each node are in order of increasing dimension.}
 \label{fig:normal-fixed}
 \end{figure}
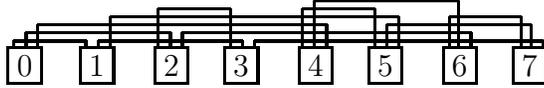

If we require that each node has its connections in order of
dimensions $1$, $2$, \ldots~$\lgN$, we cannot achieve a routing
in $m(N)$ tracks when $N>2$; Figure~\ref{fig:normal-fixed} with 6
tracks shows the best layout of a row when $N=8$.  Even with this
fixed order of connections, however, the density (and therefore the
number of tracks) is just $m(N)+1$ for $N>2$.  Our approach to
obtaining this stronger result also produces a characterization of
{\em all\/} locations where the density is maximized.  We encapsulate
these results in the following two Theorems.

\begin{thm}
 \label{thm:i's-maximizing-S}
 The values of $i$ in binary for which $S(i,N)$ is maximized are those
obtained as follows.  Starting from the leftmost bit
of $i$ and moving right, choose pairs of bits to be 01 or 10 except
that when $\lgN$ is even, the last pair may be 11.  When
$\lgN$ is odd, the 1 remaining bit is set to 1.
 \end{thm}

\begin{pf}
 Considering the number $i$ represented in binary, define
 $b(i,j)$ to be the bit in the $j$-th position from the right ($1\leq
j\leq\lgN$), and define $e(i,j)$ to be the excess of 1's over
0's in bit positions greater than $j$ (i.e., the number of 1's minus
the number of 0's in the relevant portion of $i$'s representation).
 Also, let $r$ denote the number of consecutive 0's at the right end
of $i$'s representation.
 (Using the notation $0^r$ to represent a string of $r$ 0's, note that
with $i$ of the form $X10^r$, $i-1$ is $X01^r$, and $-i$ is
$\overline{X}10^r$, where $\overline{X}$ is the bitwise complement of
$X$.)
 Starting from the definitions of $S(i,N)$ and $f(i,k)$ in
Equations~\ref{eqn:S-i-n} and~\ref{eqn:f-i-k}, we
 \extended{can express $S(i,N)$ as\small}{see that}
 \newcommand{\compactsumcore}[2]{(1-2b(i,j))e(#1,#2)}
 \newcommand{\widesumcore}[2]{\left\{\begin{array}{ll}
                    e(#1,#2) & \mbox{if $b(i,j)=0$} \\
                    -e(#1,#2) & \mbox{if $b(i,j)=1$}
                    \end{array}\right.}
 \extended{\newcommand{\sumcore}{\compactsumcore}}{\newcommand{\sumcore}{\widesumcore}}
 \begin{eqnarray}
 \extended{&&\hspace*{-2em}}{S(i,N) & = &}
         \sum_{k=1}^{\lgN}i(1-2b(i-1,k))\bmod2^k \nonumber \\
   & = & \sum_{k=1}^{\lgN}\sum_{j=1}^{k}b(i(1-2b(i-1,k)),j)\cdot2^{j-1} \nonumber \\
 \extended{}{
   & = & \sum_{j=1}^{\lgN}\sum_{k=j}^{\lgN}b(i(1-2b(i-1,k)),j)\cdot2^{j-1} \nonumber \\}
   & = & \sum_{j=1}^{\lgN}\sum_{k=j}^{\lgN}2^{j-1}\left\{
         \begin{array}{ll}
         b(i,j) & \mbox{if $b(i-1,k)=0$} \\
         b(-i,j) & \mbox{if $b(i-1,k)=1$}
         \end{array}
         \right. \nonumber \\
   & = & \sum_{j=1}^{\lgN}2^{j-1}\left[\frac{b(-i,j)+b(i,j)}{2}(\lgN-j+1)
                                         +\frac{b(-i,j)-b(i,j)}{2}e(i-1,j-1)\right] \nonumber \\
   & = & \sum_{j=1}^{r}0 + 2^{r}(\lgN-r)
         + \sum_{j=r+2}^{\lgN}2^{j-2}\left[\lgN-j+1
         + \sumcore{i-1}{j-1}
           \right] \nonumber \\
   & = & 2^{r}(\lgN-r)+\lgN\sum_{j=r+2}^{\lgN}2^{j-2}
         - \sum_{j=r+2}^{\lgN}j2^{j-2}
         + \sum_{j=r+2}^{\lgN}2^{j-2}\sumcore{i}{j} \nonumber \\
   & = & \half N+\sum_{j=r+2}^{\lgN}2^{j-2}\sumcore{i}{j} \nonumber \\
   & = & \half N-\sum_{j=1}^{r+1}2^{j-2}\sumcore{i}{j}
         + \sum_{j=1}^{\lgN}2^{j-2}\sumcore{i}{j} \nonumber \\
   & = & \half N-\sum_{j=1}^{r}2^{j-2}(e(i,0)+j)+2^{r-1}(e(i,0)+r-1)
         + \sum_{j=1}^{\lgN}2^{j-2}\sumcore{i}{j} \nonumber \\
   & = & \half(e(i,0)+N-1) +
 \sum_{j=1}^{\lgN-1}2^{j-2}\cdot\widesumcore{i}{j}
 \ .
 \label{eqn:S-i-n-revised}
 \end{eqnarray}
 \normalsize
 From this expression, we can see that $S(i,N)$ is maximized by
setting pairs of bits greedily from the left end of $i$'s
representation, except for a slight variation when $j$ becomes small,
as in the theorem statement.  (It is also easy to check that this
maximum equals $m(N)$ of Equation~\ref{eqn:mN}.)
 \qed
 \end{pf}

Now we proceed to analyze the maximum density in a row of the layout
when it is required that each node has its connections in order of
dimensions $1$, $2$, \ldots~$\lgN$.  We define $T(i,p,N)$ to be
the number of wires crossing a cutline just to the right of the $p$-th
terminal position on a node in column $i-1$ for $1\leq
p\leq\lgN$ (so $T(i,\lgN,N)=S(i,N)$).

\begin{thm}
 For $N>2$, the maximum value of $T(i,p,N)$ over all $i$ and $p$ is
$m(N)+1$ and is realized at an $i$ for which $S(i,N)=m(N)$.
 \end{thm}

\begin{pf}
 We can express $T(i,p,N)$ in terms of $S(i,N)$ by using the notation
defined at the beginning of the proof of
Theorem~\ref{thm:i's-maximizing-S}; specifically, $T(i,p,N) =
S(i,N)+e(i-1,p)$.  The term $e(i-1,p)$ can be reexpressed in terms of
$e(i,p)$ based on the value of $r$ defined above.  For $p>r$, we have
$e(i-1,p)=e(i,p)$.  For $p\leq r$, we have $e(i-1,p)=e(i,p)+2(r-p-1)$.

When $r=0$, we know $p>r$, and we see that the strategy for choosing
$i$ described in Theorem~\ref{thm:i's-maximizing-S} remains optimal,
since the $e(i,p)$ term is small compared to $2^{j}$ for most values
of $j$ in Equation~\ref{eqn:S-i-n-revised}.  With such an $i$, the
largest $e(i,p)$ we can achieve is $1$ (if at least one of the pairs
of bits under the strategy of Theorem~\ref{thm:i's-maximizing-S} is
$10$ or $11$).

When $r=1$, the situation is essentially the same as for $r=0$, except
that we must choose $p>1$ to maximize $e(i-1,p)$.  We still must
choose an $i$ that maximizes $S(i,N)$, and $e(i-1,p)$ will be at most
$1$.

Choosing $r\geq2$ contradicts choosing $i$ to maximize $S(i,N)$, and
the deficit in the value of $S(i,N)$ cannot be recouped through the
term $e(i-1,p)$.  (For $r=2$, $e(i-1,p)$ cannot exceed $e(i,p)$, while
increasing values of $r$ cause increasing deterioration in the value
of $S(i,N)$.)
 \qed
 \end{pf}

Note that this result is not an idiosyncrasy of the particular
ordering chosen for the terminals on each node.  Rather, because of
the symmetry in the layout, it is apparent than any ordering that is
the same for all nodes leads to $m(N)+1$ tracks; an ordering that
reduces $T(i,p,N)$ where it exceeds $m(N)$ will make a corresponding
increase from $m(N)$ to $m(N)+1$ in another position.

 \label{sec:tracks}

\section{Alternative layouts}
\begin{figure}
 \centering
 \setlength{\unitlength}{0.00083300in}%
\begingroup\makeatletter\ifx\SetFigFont\undefined
\def\x#1#2#3#4#5#6#7\relax{\def\x{#1#2#3#4#5#6}}%
\expandafter\x\fmtname xxxxxx\relax \def\y{splain}%
\ifx\x\y   
\gdef\SetFigFont#1#2#3{%
  \ifnum #1<17\tiny\else \ifnum #1<20\small\else
  \ifnum #1<24\normalsize\else \ifnum #1<29\large\else
  \ifnum #1<34\Large\else \ifnum #1<41\LARGE\else
     \huge\fi\fi\fi\fi\fi\fi
  \csname #3\endcsname}%
\else
\gdef\SetFigFont#1#2#3{\begingroup
  \count@#1\relax \ifnum 25<\count@\count@25\fi
  \def\x{\endgroup\@setsize\SetFigFont{#2pt}}%
  \expandafter\x
    \csname \romannumeral\the\count@ pt\expandafter\endcsname
    \csname @\romannumeral\the\count@ pt\endcsname
  \csname #3\endcsname}%
\fi
\fi\endgroup
\begin{picture}(3639,499)(139,242)
\thicklines
\put(151,254){\makebox(6.6667,10.0000){\SetFigFont{10}{12}{rm}.}}
\put(1051,479){\line( 0, 1){ 50}}
\put(1051,529){\line( 1, 0){300}}
\put(1351,529){\line( 0,-1){ 50}}
\put(1771,269){\framebox(195,210){}}
\put(2236,269){\framebox(195,210){}}
\put(2671,269){\framebox(195,210){}}
\put(3136,269){\framebox(195,210){}}
\put(3571,269){\framebox(195,210){}}
\put(871,269){\framebox(195,210){}}
\put(421,269){\framebox(195,210){}}
\put(1336,269){\framebox(195,210){}}
\put(601,479){\line( 0, 1){ 50}}
\put(601,529){\line( 1, 0){300}}
\put(901,529){\line( 0,-1){ 50}}
\put(976,479){\line( 0, 1){200}}
\put(976,679){\line( 1, 0){2250}}
\put(3226,679){\line( 0,-1){200}}
\put(526,479){\line( 0, 1){100}}
\put(526,579){\line( 1, 0){1350}}
\put(1876,579){\line( 0,-1){100}}
\put(2401,479){\line( 0, 1){ 50}}
\put(2401,529){\line( 1, 0){300}}
\put(2701,529){\line( 0,-1){ 50}}
\put(451,479){\line( 0, 1){150}}
\put(451,629){\line( 1, 0){3315}}
\put(3766,629){\line( 0,-1){150}}
\put(2326,479){\line( 0, 1){100}}
\put(2326,579){\line( 1, 0){1365}}
\put(3691,579){\line( 0,-1){100}}
\put(2851,479){\line( 0, 1){ 50}}
\put(2851,529){\line( 1, 0){315}}
\put(3166,529){\line( 0,-1){ 50}}
\put(1501,479){\line( 0, 1){ 50}}
\put(1501,529){\line( 1, 0){300}}
\put(1801,529){\line( 0,-1){ 50}}
\put(1951,479){\line( 0, 1){ 50}}
\put(1951,529){\line( 1, 0){300}}
\put(2251,529){\line( 0,-1){ 50}}
\put(3301,479){\line( 0, 1){ 50}}
\put(3301,529){\line( 1, 0){315}}
\put(3616,529){\line( 0,-1){ 50}}
\put(1426,479){\line( 0, 1){250}}
\put(1426,729){\line( 1, 0){1350}}
\put(2776,729){\line( 0,-1){250}}
\put(518,314){\makebox(0,0)[b]{\smash{\SetFigFont{12}{14.4}{rm}0}}}
\put(968,314){\makebox(0,0)[b]{\smash{\SetFigFont{12}{14.4}{rm}1}}}
\put(1433,314){\makebox(0,0)[b]{\smash{\SetFigFont{12}{14.4}{rm}3}}}
\put(1868,314){\makebox(0,0)[b]{\smash{\SetFigFont{12}{14.4}{rm}2}}}
\put(2333,314){\makebox(0,0)[b]{\smash{\SetFigFont{12}{14.4}{rm}6}}}
\put(2768,314){\makebox(0,0)[b]{\smash{\SetFigFont{12}{14.4}{rm}7}}}
\put(3233,314){\makebox(0,0)[b]{\smash{\SetFigFont{12}{14.4}{rm}5}}}
\put(3668,314){\makebox(0,0)[b]{\smash{\SetFigFont{12}{14.4}{rm}4}}}
\end{picture}
 \caption{The top row of a gray code derived layout for $N=8$.}
 \label{fig:gray-float}
 \end{figure}
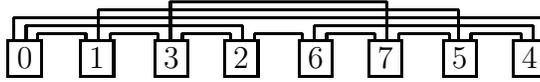

 Another frequently considered method of mapping hypercube nodes to a
regular grid is to use a gray code derived layout.  The numbering of
nodes in the top row of a gray code layout for a 64-node hypercube is
illustrated in Figure~\ref{fig:gray-float}.  (Here we have not
required the terminals on each node to be in dimension order.)  Ranade
and Johnsson~\cite{RanadeJ1987} noted that the area and maximum wire
length for the normal layout and the gray code layout are the same up
to a constant factor.  In fact, the arguments of
Sections~\ref{sec:density} and~\ref{sec:tracks} can be extended to
show that the maximum wire density and number of wiring tracks
required per row is exactly the same for the gray code layout as for
the normal layout, including a one track penalty when the nodes are
identical.  It is also easy to show that the total (horizontal) wire
length per row is the same (in terms of the number of columns
spanned).  The maximum (horizontal) wire length in a row of the normal
layout, however, is essentially half as large as for the gray code
layout.

The results of Harper~\cite{Harper1964,Harper1966},
Nakano~\cite{Nakano1994}, and~Bezrukov \etal~\cite{BezrukovCHRS2000}
show that the normal layout minimizes total wire length and
intercolumn wire density, while a different layout minimizes maximum
wire length.  Bezrukov \etal\ also consider two new cost measures for
embeddings of hypercubes into grids based on the frequent use of
normal algorithms~\cite{BezrukovCHRS1998}.

 \label{sec:alt}

\bibliographystyle{hplain}
\bibliography{sources}

\begin{thebibliography}{10}

\bibitem{AbrahamP1991}
Seth Abraham and Krishnan Padmanabhan.
\newblock Performance of multicomputer networks under pin-out constraints.
\newblock {\em Journal of Parallel and Distributed Computing}, pages 237--248,
  December 1991.

\bibitem{BezrukovCHRS1998}
S.~L. Bezrukov, J.~D. Chavez, L.~H. Harper, M.~R\"{o}ttger, and U.-P.
  Schroeder.
\newblock Embedding of hypercubes into grids.
\newblock In {\em MFCS '98}, pages 693--701. Springer-Verlag, 1998.
\newblock Lecture Notes in Computer Science 1450.

\bibitem{BezrukovCHRS2000}
S.~L. Bezrukov, J.~D. Chavez, L.~H. Harper, M.~R\"{o}ttger, and U.-P.
  Schroeder.
\newblock The congestion of $n$-cube layout on a rectangular grid.
\newblock {\em Discrete Mathematics}, 213:13--19, 2000.

\bibitem{ChenAB1993}
Chienhua Chen, Dharma~P. Agrawal, and J.~Richard Burke.
\newblock {dBCube}: A new class of hierarchical multiprocessor interconnection
  networks with area efficient layout.
\newblock {\em IEEE Trans.\ Parallel and Distributed Systems},
  4(12):1332--1344, December 1993.

\bibitem{Dally1990}
William~J. Dally.
\newblock Performance analysis of $k$-ary $n$-cube interconnection networks.
\newblock {\em IEEE Trans.\ Computers}, 39(6):775--785, June 1990.

\bibitem{GreenbergG1996ECA}
Ronald~I. Greenberg and Lee Guan.
\newblock An empirical comparison of area-universal and other parallel
  computing networks.
\newblock In {\em Proceedings of the ISCA 9th International Conference on
  Parallel and Distributed Computing Systems}, pages 260--267, September 1996.

\bibitem{GreenbergG1996ECN}
Ronald~I. Greenberg and Lee Guan.
\newblock An empirical comparison of networks and routing strategies for
  parallel computation.
\newblock In {\em Proceedings of the Eighth IASTED International Conference
  Parallel and Distributed Computing and Systems}, pages 265--269, Chicago,
  October 1996.

\bibitem{Harper1964}
L.~H. Harper.
\newblock Optimal assignments of numbers to vertices.
\newblock {\em Journal of the Society for Industrial and Applied Mathematics},
  12(1):131--135, March 1964.

\bibitem{Harper1966}
L.~H. Harper.
\newblock Optimal numberings and isoperimetric problems on graphs.
\newblock {\em Journal of Combinatorial Theory}, 1:385--393, 1966.

\bibitem{HashimotoS1971}
Akihiro Hashimoto and James Stevens.
\newblock Wire routing by optimizing channel assignment within large apertures.
\newblock In {\em \dac{8}}, pages 155--169. IEEE Computer Society Press, 1971.

\bibitem{Leighton1992}
F.~Thomson Leighton.
\newblock {\em Introduction to Parallel Algorithms and Architectures: Arrays
  $\cdot$ Trees $\cdot$ Hypercubes}.
\newblock Morgan Kaufmann, 1992.

\bibitem{Lengauer1990}
Thomas Lengauer.
\newblock {\em Combinatorial Algorithms for Integrated Circuit Layout}.
\newblock John Wiley, 1990.

\bibitem{Nakano1994}
Koji Nakano.
\newblock Linear layouts of generalized hypercubes.
\newblock In {\em Proceedings of the 19th International Workshop on
  Graph-Theoretic Concepts in Computer Science (WG '93)}, pages 364--365.
  Springer-Verlag, 1994.

\bibitem{RanadeJ1987}
Abhiram~G. Ranade and S.~Lennart Johnsson.
\newblock The communication efficiency of meshes, boolean cubes and cube
  connected cycles for wafer scale integration.
\newblock In {\em \icpp{1987}}, pages 479--482, 1987.

\bibitem{Thompson1979}
C.~D. Thompson.
\newblock Area-time complexity for {VLSI}.
\newblock In {\em \stoc{11}}, pages 81--88. ACM Press, 1979.

\bibitem{Thompson1980}
C.~D. Thompson.
\newblock {\em A Complexity Theory for {VLSI}}.
\newblock PhD thesis, Department of Computer Science, Carnegie-Mellon
  University, 1980.

\end{thebibliography}

\end{document}